\documentclass[aps,prb,twocolumn,nopacs,floats,superscriptaddress]{revtex4-1}
\usepackage{graphicx} % Required for inserting images
\usepackage{dcolumn}
\usepackage{bm}
\usepackage{amsmath}
\usepackage{color}
\UseRawInputEncoding
\def\he4{$^4$He}
\def\hel3{$^3$He}
\def\Am3{\AA$^{-3}$}
\def\beq{\begin{equation}}
\def\eeq{\end{equation}}

\newcommand{\pt}{{\partial}}

\newcommand{\cL}{{\mathcal L}}

\newcommand{\be}{\begin{equation}}
\newcommand{\ee}{\end{equation}}
\newcommand{\bea}{\begin{eqnarray}}
\newcommand{\eea}{\end{eqnarray}}
\newcommand{\bse}{\begin{subequations}}
\newcommand{\ese}{\end{subequations}}
\def\rf#1{(\ref{#1})}

\begin{document}

\title{Transverse Quantum Fluids}
\author{Anatoly Kuklov}
\affiliation{Department of Physics \& Astronomy, College of Staten Island and the Graduate Center of
CUNY, Staten Island, NY 10314}

\author{Nikolay Prokof'ev}
\affiliation{Department of Physics, University of Massachusetts, Amherst, MA 01003, USA}

\author{Leo Radzihovsky}
\affiliation{ Department of Physics and Center for Theory of Quantum Matter, University of Colorado, Boulder, CO 80309}

\author{Boris Svistunov}
\affiliation{Department of Physics, University of Massachusetts, Amherst, MA 01003, USA}
\affiliation{Wilczek Quantum Center, School of Physics and Astronomy and T. D. Lee Institute, Shanghai Jiao Tong University, Shanghai 200240, China}

\date{\today}

\begin{abstract}
  Motivated by the remarkable properties of superfluid edge dislocations
  in \he4, we discuss a broad class of quantum systems---boundaries in
  phase separated lattice states, magnetic domain walls, and ensembles
  of Luttinger liquids---that can be classified as Transverse Quantum
  Fluids (TQF). After introducing the general idea of a TQF, we focus on
  a coupled array of Luttinger liquids forming an incoherent TQF. This
  state is a long-range ordered quasi-one-dimensional superfluid,
  topologically protected against quantum phase slips by the tight-binding
  of instanton dipoles, that has no coherent quasi-particle
  excitations at low energies. An incoherent TQF is a striking example of
  the irrelevance of the Landau quasiparticle criterion for
  superfluidity in systems that lack Galilean invariance. We detail
  its phenomenology, to motivate a number of experimental studies in
  condensed matter and cold atomic systems.
\end{abstract}
%\pacs{ 67.80.bd, 67.80.dj, 67.80.-s, 67.80.B-}
% 67.80.bd Superfluidity in solid 4He, supersolid 4He
% 67.80.dj Defects, impurities, and diffusion
% 67.80.-s Quantum solids
% 67.80.B- Solid 4He

\maketitle

{\em Introduction and motivation.} In the last two decades, there has been much attention on systems
where nontrivial gapless physics is confined to a surface of a bulk
material, with such prominent examples as topological insulators
\cite{HasanKaneTI} and their symmetry-protected topological
generalizations \cite{SenthilSPT}. In a broader sense, these are
examples of a large class of low-dimensional systems in which the
surrounding bulk host is fundamentally important for understanding
their properties. Such systems include randomly pinned surface
Goldstone modes \cite{surfaceLRnematic, surfaceLRsmectic},
an edge state of a confinement-free quantum Hall droplet \cite{KunYangEdge},
a superfluid interface between two checkerboard-solid domains \cite{Burovskii},
as well as open spin/particle chains coupled to a dissipative bath \cite{NayakOpenLL,Pankov,Werner,Caza,Lobo,Cai,Lode,Weber,Danu,Martin}.

A particularly interesting system studied recently is an edge
dislocation with a superfluid core \cite{sclimb,edge2022,TQFprl} in
solid \he4 where the low-temperature motion of the dislocation
transverse to its core and Burgers vector (the so-called climb)
requires boson number flow onto the core.  The corresponding
superfluid state of the edge dislocation has been dubbed as a transverse
quantum fluid (TQF), a new one-dimensional state of bosons
qualitatively distinct from a Luttinger liquid (LL) with: (i) a
quadratic spectrum of excitations, (ii) off-diagonal long-range order
at $T=0$, (iii) exponential dependence of the phase slip probability
on the inverse flow velocity, and    (iv) nonapplicability of Landau criterion.
The key ingredient responsible for all these features is the translational invariance with respect to the core motion in the climb direction, 
necessarily accompanied by atom transfer to the \he4 crystal bulk.  Most importantly,  
this implies  {\it infinite compressibility}, which is solely responsible for properties (i)-(iii).

\begin{figure}[!htb]
%\vskip-8mm
\includegraphics[width=0.7 \columnwidth]{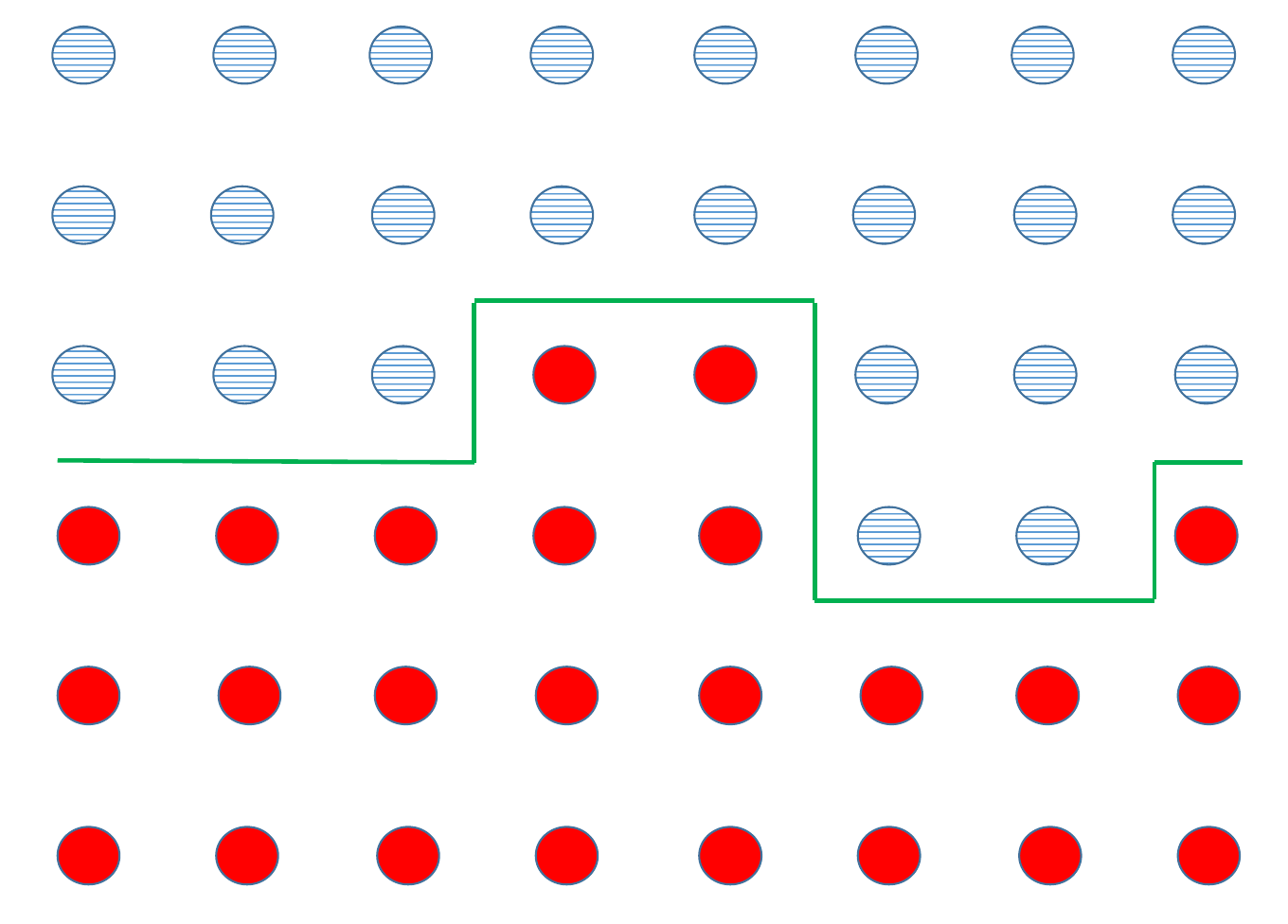}
%	\vskip-8mm
\caption{ A sketch of models A, B, C, where the domain
  line (green) is proposed to realize a TQF.  In cases A and C,
  the solid (red) circles represent bosons forming a Mott insulating
  phase, while the rest of the sites shown by patterned circles remain
  empty (model A), or represent particles of the second bosonic
  component also forming a Mott insulating phase when the mixture is
  immiscible (model C).  Particle motion by exchange along the phase
  separation (green) line---the domain wall---corresponds to the
  deformation of the domain wall in direct analogy with atom
  redistribution along the edge dislocation core under its climb. For
  model B, different circles represent opposite orientations of
  magnetization across the Bloch domain wall \cite{Bulaevski}. }
	\label{fig1}
\end{figure}

\begin{figure}[!htb]
%\vskip-8mm
\includegraphics[width=0.9 \columnwidth]{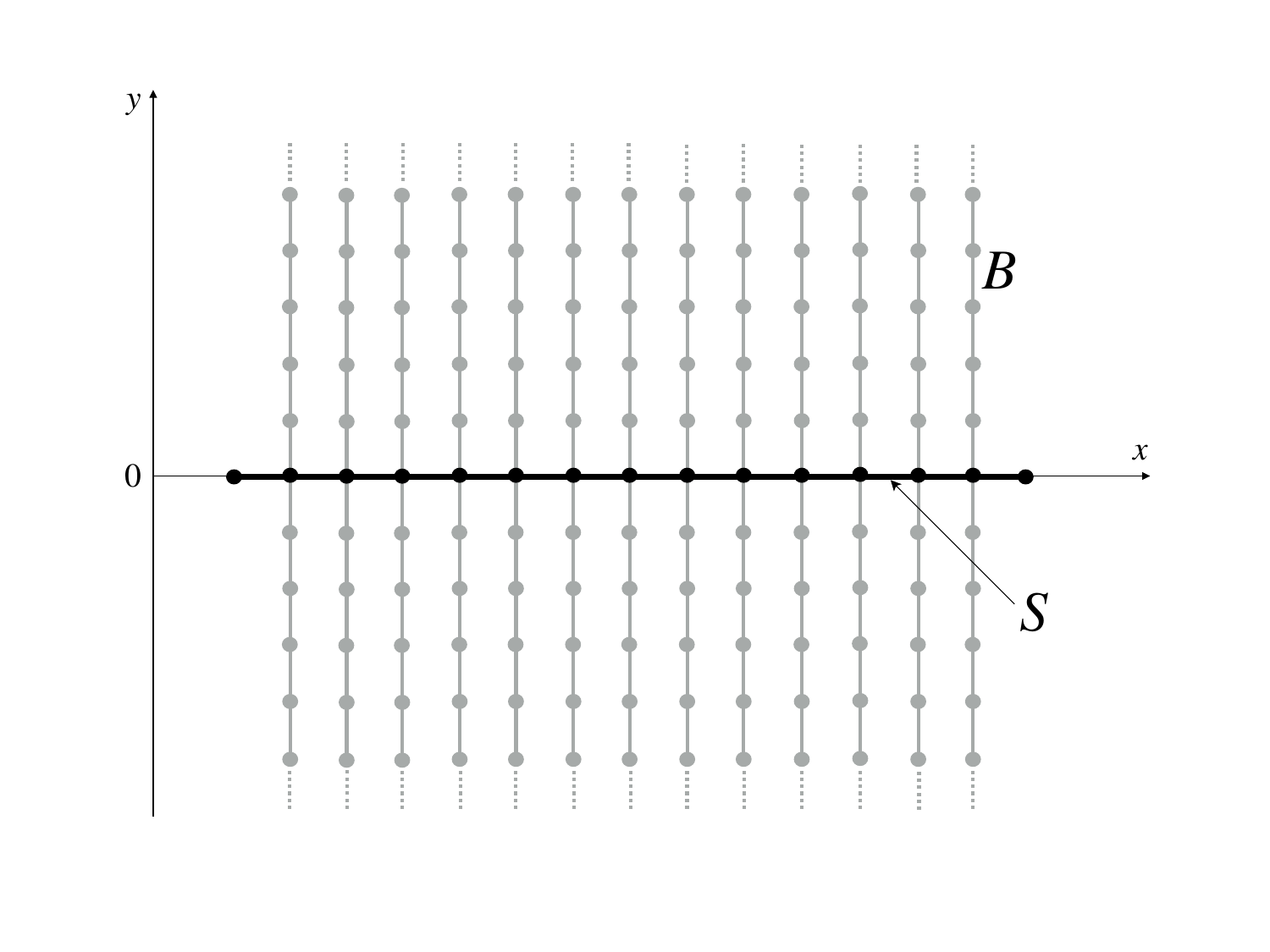}
%	\vskip-8mm
\caption{A proposed realization of the incoherent TQF in a bosonic
  system on a square lattice, where vertical ``bath'' LLs are coupled
  to the horizontal ``system'' LL at $y=0$, but are otherwise
  decoupled.  The setup is similar to the one proposed in
  Ref.~\cite{Lode} and can also be viewed as a coupled chain of
  Kane-Fisher dots \cite{KaneFisher}.  }
\label{fig2}
\end{figure}

 In this Letter, we argue that similar---infinite-compressibility-driven---phenomenology emerges in other
interesting examples and discuss four specific systems that we predict
to host a TQF.  These
are (A) a self-bound droplet of hard core bosons on a two-dimensional
(2D) lattice, (B) a Bloch domain wall in an easy-axis
  ferromagnet, (C) a phase separated state of two-component bosonic Mott
  insulators with the boundary in the counter-superfluid phase
  \cite{SCF} (or in a phase of a two-component superfluid) on a 2D lattice, as
  illustrated in Fig.~\ref{fig1}, and (D) a  1D bosonic
liquid Josephson-coupled to a collection of transverse LLs, which are
otherwise decoupled from each other---a setup in Fig.~\ref{fig2},
similar to the one considered in Ref.~\cite{Lode}, and also related closely to a number of other setups and models discussed in the past \cite{Pankov,Werner,Caza,Lobo}.
While A, B, and C are naturally forming edge systems that share the same low-energy description with the superclimbing
dislocation \cite{sclimb,TQFprl},
system D is rather an ``engineered" state with infinite compressibility distinguished by its lack of
well-defined elementary excitations altogether;  yet, it is a superfluid
by other hallmark properties---(ii), (iii), and (iv) above.

In the context of these TQF systems, we note that historically the
stability of a homogeneous superflow with velocity $v$ was linked to
the Landau criterion $v < \min{\epsilon(k)/k}$, where $\epsilon (k)$
is the dispersion of elementary quasi-particle excitations.  In the
absence of Galilean invariance, this criterion can be strongly
violated and, in particular, does not apply to models A, B, and C,
which feature $\epsilon (k) \propto k^2$. This observation is even
more striking for model D lacking well-defined elementary excitations,
even precluding the applicability of the Landau criterion.

We begin by introducing the TQF in the A, B, and C systems as sketched
in Fig.~\ref{fig1}.  At the level of low-energy description, these three systems are equivalent.
%%%%%%%%%%%%%%%%%%%%%%%%%%%%%%%%%%%%%%%%%%%%%%%%%%%%
%%%%%%%%%%%%%%%%%%%%%%%%%%%%%%%%%%%%%%%%%%%%%%%%%
 The boundary of a bulk Mott insulator (model A) can
  be superfluid, provided the bulk is close to the superfluid
  transition. A domain wall in a magnet ( model B) formed between
  domains with opposite easy-axis magnetization, can undergo a
  transition from the Ising-type to the Bloch-type characterized by
  the easy-plane magnetization, free to choose any direction.  The
  description of the classical transition between these two phases in
  a 3D ferromagnet was introduced more than 60 years ago in
  Ref.~\cite{Bulaevski}. The Bloch-type domain walls also occur in the
  liquid $^3$He-A phase \cite{sasaki}.

Model C maps onto a ferromagnet, as described in Refs.~\cite{SCF,Demler}, thus making its domain wall equivalent to that in model B. At the microscopic level, the easy-axis magnetization in model C describes counterflow superfluidity of the components.   For each of these proposed realizations of
  TQF the domain line must be in the quantum-rough state.
This state occurs naturally in the vicinity of the phase separation transition or close to the transition from the Ising to the Bloch-type domain wall. 
 Conversely,
  if the domain line is even weakly pinned to be quantum smooth, at
  sufficiently long scales, the TQF crosses over to a conventional LL
  state.

The regime of the self-pinned lines has been studied in numerous publications (see
Refs.~\cite{DWLL1,DWLL2, DWLL3} and references therein).  In contrast,
a quantum-{\em rough} Bloch domain wall is a 1D quantum fluid that is
qualitatively distinct from a LL, with a low-energy description given
by a TQF Hamiltonian \cite{TQFprl}
\begin{equation}
 H[\phi,n] = \int \left[ \, \frac{\chi}{2} (\partial_x n)^2 + \frac{n_s}{2} (\partial_x \phi )^2 \, \right] \, dx ,
\label{HTQF}
\end{equation}
expressed in terms of the superfluid phase (or, equivalently, the angular orientation of the easy-axis magnetization) $\phi (x)$ and the
canonically conjugate 1D projected density $n(x)$ proportional to
the domain wall transverse (vertical, $y$ in Fig.~\ref{fig1})
displacement.  In \rf{HTQF}, $n_s$ is the superfluid/spin stiffness,
and $\chi$ is the domain wall line tension (energy per unit length).
%that is, $(\partial_x n)^2/2$ describes an elongation of the wall due
%to its smooth deformation (in the leading order).
For edge dislocations, $\chi$ is fixed by the lattice shear modulus;
for a domain wall in models A, B, and C, the $\chi$ stiffness can be
tuned by proximity of the bulk material to the corresponding critical
point.  The key feature distinguishing TQF from LL is its divergent
compressibility, i.e., absence of the leading interaction term
$\sim n^2$ in (\ref{HTQF}). This leads to (i) a quadratic dispersion,
$\epsilon_k \, = \, \sqrt{\chi n_s} \, k^2$, for elementary
excitations with linear momentum $k$, (ii) off-diagonal long-range
order, and (iii) exponential suppression of the phase slip events
\cite{TQFprl}.

An intriguing aspect of a 1D superfluid interface between two
insulating ground states is the emergent link between two seemingly
unrelated properties---superfluidity and roughness.  Depending on the
type of the insulating state(s), superfluidity and roughness can be
mutually exclusive or inevitably linked.  The latter situation takes
place in the vicinity of the superfluid-checkerboard solid quantum
critical point when the deconfinement of spinons converts the smooth insulating
domain wall into a rough superfluid \cite{Burovskii}.  %In a TQF-type of an interface, superfluidity ultimately leads to a smooth boundary in the thermodynamic limit \cite{TQFprl}.

Finite compressibility $\kappa$ may be induced in TQF (and thus transforming it into LL) either by an
external potential or through lattice pinning, as discussed in
Ref.~\cite{TQFprl}. In the presence of thereby generated
interaction energy $\kappa^{-1} n^2$ in (\ref{HTQF}), on a length scale
beyond $\xi = \sqrt{\chi \kappa}$ the system exhibits a TQF-to-LL
crossover, with the familiar low-energy linear spectrum and
algebraically decaying off-diagonal correlations. This crossover can
be suppressed if the domain wall is incommensurately tilted (by
pinning the end points) and has a finite concentration of kinks larger
than $\xi^{-1}$, as discussed in Ref.~\cite{Max}.

{\it Incoherent superfluid}. We now focus on a qualitatively
distinct and most interesting model D---a particularly simple microscopic representative of a family of long-ranged ordered dissipative models \cite{Pankov,Werner,Caza,Lobo}---where infinite compressibility
and TQF properties emerge due to Josephson coupling to the
``transverse" bulk array of LLs, rather than transverse domain wall
fluctuations.  As illustrated in Fig.~\ref{fig2}, this model can be
realized either with cold atoms or superconducting wires, and consists
of a ``system'' LL (running along $x$), with strong Josephson links to
a transverse array of identical and independent ``bath'' LLs labeled
by index $i$. All LLs are taken to be in a superfluid regime,
characterized by a Luttinger parameter $K > 1$.
 Coupling between the bath LL must be negligible to avoid
 a global superfluid phase or gapped bulk insulator.
[This setup was proposed and numerically simulated in Ref.~\cite{Lode},
where, however, the TQF superfluidity of the system along the $x$
direction was neither recognized nor explored.]

For model D, the Euclidean action,
\[
S[\phi(\tau,x),\varphi_i(\tau,y)]= \int d\tau dx (\cL_s + \cL_{\rm
  int} ) + \int d\tau dy \cL_b ,
  \]
 is based on three Lagrangian densities,
\be \cL_s={\kappa_s\over2}(\pt_\tau\phi)^2 +
{n_s\over2}(\pt_x\phi)^2\, ,
\label{Ls}
\ee
\be
\cL_{\rm int} = \frac{g}{2}\sum_i \left[\phi(x) - \varphi_i(y=0)
  \right]^2\delta(x-x_i) \, ,
\label{Lint}
\ee
\be
\cL_b=\sum_i\left[{\kappa_b\over2}(\pt_\tau\varphi_i)^2 +
{n_b\over2}(\pt_y\varphi_i)^2\right] \, ,
\label{Lb}
\ee
where $\phi\equiv \phi(x,\tau)$ and
$\varphi_i \equiv \varphi_i(y,\tau)$ are the local superfluid phases
of the system and bath LLs, respectively, $g>0$, and
$\kappa_s,\, \kappa_b$ and $n_s, \, n_b$ are, respectively,
compressibilities and superfluid stiffnesses. In (\ref{Lint}) the
Josephson coupling $-g\cos[\phi(x,\tau) - \varphi_i(y=0,\tau)]$ has
been approximated by the quadratic form, valid for strong coupling $g$
in the LLs' superfluid regime, $K>1$, where $g$ flows to
infinity \cite{KaneFisher}. Since we are interested in the
long-wavelength limit, we go to the continuum by replacing $\sum_i$
with $\int dx/a$ and $\varphi_i(y)$ with $\varphi(x,y)$, and take the
lattice constant $a$ as the unit of length, or, equivalently absorbing
it into the definition of model parameters.

The bath phase $\varphi$ can be straightforwardly integrated out of
the quadratic Lagrangian, decoupled in the momentum space,
\be
S={1\over2}\! \sum_{\omega,k_x} \! \left[(\kappa_s\omega^2 \! + \! n_sk_x^2) |\phi_{\omega,k_x}|^2 \! + g |\phi_{\omega,k_x} \!\! - \tilde{\varphi}_{\omega,k_x}|^2  \right] + S_b,
\label{S3}
\ee
\be
S_b={1\over2}\sum_{\omega, k_x,k_y} \left[\kappa_b\omega^2+ n_b k_y^2 \right]|\varphi_{\omega,k_x,k_y}|^2,
\label{S4}
\ee
with
$\tilde{\varphi}_{\omega, k_x} = L_y^{-1/2} \sum_{k_y}
\varphi_{\omega,k_x,k_y}$.  We thereby get the effective 1D system
action
%({\color{red}after taking the limit $g\to \infty$):}
%
%\be
%S= {1\over2}\sum_{\omega,k_x}
%\left[\kappa_s\omega^2 + K_b |\omega|+n_s k_x^2  \right]|\phi_{\omega,k_x}|^2,
%\label{theta}
%\ee
%
%
\be
S= {1\over2}\sum_{\omega,k_x}
\left[(\kappa_s\omega^2 + n_s k_x^2) + \frac {gK_b |\omega| }{g + K_b
    |\omega|} \right]|\phi_{\omega,k_x}|^2,
\label{theta}
\ee
with $K_b\equiv 2\sqrt{n_b \kappa_b}$ and $g \to \infty$.  We dub the low-energy
long-wavelength limit of this action as the ``incoherent TQF'' (iTQF),
\be
S_{\rm iTQF} = {1\over2}\sum_{\omega,k_x} \left[K_b|\omega|
  + n_sk_x^2 \right]|\phi_{\omega,k_x}|^2 ,
\label{theta2}
\ee
 where $\omega \ll \omega_b=K_b/\kappa_s$.
With Wick's rotation to the real time/frequency this action describes
diffusive dynamics of system's superfluid phase, with
$\omega = - {\rm i} Dk_x^2$ with $D=n_s/K_b$.  The corresponding
real-time action can be equivalently obtained using Feynman-Vernon and
Schwinger-Keldysh double-time contour methods \cite{FZcosine,
  FeynmanVernon, Kamenev}.

Next, we show that despite lacking well-defined elementary excitations
and being characterized by diffusive dynamics, iTQF exhibits a 1D
off-diagonal long-range order and exponentially suppressed probability
of the phase slip at a small superflow velocity $v$, i.e., it is a robust
superfluid.  This contrasts qualitatively with the ideal Bose gas and
the 1D LL, that are, respectively unconditionally and power-law
unstable at nonzero $v$.

{\it Off-diagonal long-range order}. The single-particle density
matrix at $T=0$ can be straightforwardly evaluated for the Gaussian
action (\ref{theta2}) as
\be
\langle e^{i \phi(x,0)} e^{-i \phi(0,0)}\rangle =
\exp\left[ - \int \frac{d\omega dk_x}{(2\pi)^2} \frac{1-\cos(k_x x)}{K_b |\omega| +n_s k_x^2} \right] .
\label{ODLRO}
\ee
The integral in the exponent saturates to a constant at large
separations $x$, i.e., the system is phase ordered and exhibits a
nonzero condensate fraction. In contrast, at nonzero temperature $T$
the integral over frequency is replaced with a discrete Matsubara sum
$T\sum_{\omega_n}$ and $\omega \to \omega_n=2\pi T n$, controlled by
$\omega_n = 0$ classical contribution, thereby leading to the
asymptotic exponential decay of the density matrix with exponent
$\propto - (T/2n_s) x$. This law is generic to equilibrium classical
phase fluctuations associated with 1D classical Hamiltonian density
$(\partial\phi/\partial x)^2$, insensitive to the nature of quantum
dynamics.

{\it Quantum phase slips (instantons)}. The superflow at zero
temperature decays by quantum phase slips \cite{ColemanInstantons}.
Within the leading exponential approximation, the probability of this
tunneling process can be estimated via the Euclidean action associated
with the unbinding of instantons carrying opposite topological
``charges,'' $\pm q$ (integer multiples of $2\pi$), which measure the
phase winding around singular points in space and imaginary time.  In
conventional superfluid LLs biased by a chemical potential difference
$\delta \mu$, phase slips lead to the power-law dependence (current-voltage, I-V, characteristic),
$\delta \mu \sim |v|^\alpha$ with $\alpha =2K-1$, universally
determined by the Luttinger parameter $K$ \cite{IVinLL}.

Quantum phase slips in the TQF are qualitatively different from those
in the LL because the instanton pairs in TQF are tightly
confined. This leads to the exponential dependence of the bias $\delta \mu$ on
the inverse of superflow velocity $v$.  Despite the fact that TQF
and iTQF are distinct phases, at the level of phase fluctuations (see
Eqs.~(\ref{HTQF}) and (\ref{theta2})), the dependence of the instanton
action on $v$ turns out to be the same, as we now demonstrate.

To derive the dissipation via instantons we follow the path outlined
in Ref.~\cite{TQFprl} and introduce the velocity field
$v_\mu = \pt_\mu \phi $ in the $(1+1)$-dimensional space-time
$x_\mu = (x,\tau)$. As described above, instantons have the form of
point-vortex singularities in the otherwise regular field $v_\mu$:
\begin{equation}
\pt \times v  = q(x_\mu)
  = \sum_{j} \,  q_{j} \, \delta^2 (x_\mu - x_{\mu,j}) ,
\label{instanton}
\end{equation}
where $\pt \times v \equiv\epsilon_{\mu\nu}\pt_\mu v_\nu$ is a
shorthand notation for the $(1+1)$D space-time curl of $v_\mu$ and
$x_{\mu,j}$ is the space-time position of the $j$-th instanton.  The
instanton contribution to the action (\ref{theta2}) is given by \be
S={1\over 2} \sum_{\omega, k_x} \left[ {K_b |v_\tau|^2\over |\omega |}
  + n_s |v_x|^2 - i \lambda ( ik_x v_\tau - i\omega v_x - q) \right],
\label{inst1}
\ee
where $\lambda$ is an auxiliary field enforcing the vorticity
constraint (\ref{instanton})  (for $|\omega| \ll  \omega_b)$.  The probability to find a set of
instantons with charges $\{ q_i \}$ at locations $\{ x_i \}$ is given
by the integral
\begin{equation}
  P = \int [dv_\mu][d\lambda] e^{-S }\, .
\label{Z}
\end{equation}
The Gaussian integration in the leading exponential approximation results in
\be
P \propto  e^{- \frac{1}{2}\sum_{j \ne i} V(x_{\mu,i} - x_{\mu,j})q_i q_j},
\label{Z2}
\ee
 where
\be
V(x,\tau)= \int { d\omega dk_x\over (2\pi)^2} {n_s K_b \over |\omega|( K_b|\omega| + n_s k_x^2)} [e^{i(\omega \tau + k_x x)} -1].
\label{inst}
\ee
is the instanton interaction potential.  This interaction leads to
confinement in space, $V(x,\tau=0) \sim |x|\ln |x|$ at a large separation of two instantons with opposite "charges", and weaker confinement in time,
$V(x=0,\tau) \approx -  \sqrt{2 K_b n_s \tau/\pi} $.  Apart from
the additional $\ln |x|$ with spatial separation, this behavior is
similar to that of TQF instantons found in Ref.~\cite{TQFprl}.

An imposed superflow $v$ drives instantons apart, working against
$V(x,\tau) $ via an additional contribution $2\pi n_s v \tau $ to the
instanton-pair action.  Now the corresponding action
  $S= -(2\pi)^2 \, V(x=0,\tau) - 2\pi n_s |v|\tau $ features an
extremum at $\sqrt{\tau} = \sqrt{\pi K_b n_s /2}/v$, leading to the
phase slip probability $P \propto e^{-v_1/v}$ with $v_1=\pi^2
K_b$ and predicting exponentially suppressed I-V
\begin{equation}
\delta\mu \sim e^{-v_1/v},
\label{IV}
\end{equation}
contrasting qualitatively with the power-law LL I-V.   This
  prediction holds in the asymptotic limit $v_1/|v| \gg 1$.  However,
  the optimal solution $\tau \sim 1/v^2$ breaks down at short time
  scales, for $\tau \approx 1/\omega_b= \kappa_s/K_b$, corresponding
  to a characteristic velocity $v_c \approx v_1 /\sqrt{2\pi} K_s$,
  where $K_s =\pi \sqrt{n_s\kappa_s}$ is the effective Luttinger
  parameter of the system. For $K_s \gg 1$, this velocity $v_c \ll v_1$, and we
   expect the TQF I-V (\ref{IV}) to cross over to the power-law characteristic of the LL at $|v| >v_c$.

{\it Discussion and conclusion}. While the history of supersolid
experiments in \he4 following Ref.~\cite{Kim_Chan} is rather
controversial, some of the ideas and concepts generated to understand
the phenomenon have proved to be quite generic and apply to a broad class
of physical systems. In addition to the TQF state of an edge
dislocation, similar TQF physics appears in a variety of systems:
superfluidity along the boundary between two-dimensional insulating
droplets, domain boundaries in an easy-axis ferromagnet, and in
immiscible two-component Mott insulators.  The case of iTQF is more
exotic and requires a special setup illustrated in Fig.~\ref{fig2} and
discussed in Ref.~\cite{Lode}.  However, Ref.~\cite{Lode} and numerous
studies of magnetic domain walls (see Refs.~\cite{DWLL1,DWLL2, DWLL3} and
references therein) failed to identify the iTQF and TQF properties of
these systems.

The key ingredient defining the class of TQF and iTQF states along
with their unusual properties is infinite compressibility encoded
by the effective one-dimensional field theory that we derived,
  
\be
\sigma\, =\,  \lim_{\omega \to 0} {1\over \omega^2} \lim_{k_x\to 0} {\delta^2 S[\phi_{\omega,k_x},\phi_{\omega,k_x} ^*] \over \delta \phi_{\omega,k_x}  \delta  \phi_{\omega,k_x}^*} \,=\, \infty \, .
\label{compress}
\ee
This
property is {\it sufficient} to ensure superfluid long-range order and
tight binding of instantons irrespective of other system details such
as gapped or gapless bulk excitations, the spectrum, or the very
existence of elementary excitations. The condition $\sigma=\infty$ is not only sufficient but also necessary: At $\sigma \neq \infty$, the low-energy physics of the system would correspond to that of Luttinger liquid.

To the best of our knowledge, the unusual quasi-1D superfluidity in
the domain boundaries between (and surfaces of) 2D bulk-insulating
phases was discussed only in the context of quantum Hall states in
Ref.~\cite{KunYangEdge}. However, chiral, time-reversal breaking
boundaries in this system---while featuring similar to TQF quadratic
effective field theory and thus nonacoustic dispersion of elementary
excitations---are qualitatively distinct at the fundamental
(nonlinear) level.  The hallmark of TQF---be it a ``canonical'' case
captured by superclimbing dislocations \cite{TQFprl} or by models A,
B, and C in Fig.~\ref{fig1}, or the special iTQF case---is the
tight confinement of instantons with opposite charges and the
resulting protection against the quantum phase slips.  This aspect is
absent in the chiral quantum Hall edge states, instead protected by
the gapped bulk.  Qualitatively, this situation is most close to
the surface currents in 2D superconductors. Despite their 1D
character---enforced by the Meissner effect---the currents are
protected from phase slips by the bulk order.  Moreover, it is the
physics of instantons that justifies the ``quantum'' characterization
of the TQF and iTQF states.  Otherwise, the parabolic dispersion of
TQF and the diffusive iTQF dynamics are perfectly well captured by the
corresponding classical-field theories---similar to that of the edge
currents in the quantum Hall and superconducting systems.

Finally, we emphasize that from the conceptual point of view, the iTQF
state is a striking demonstration of the conditional character of many
dogmas associated with superfluidity, such as the necessity of elementary
excitations, in general, and the ones obeying the Landau criterion in
particular, as well as the absence of long-range order in
one-dimensional quantum superfluids. Experimental and numerical
implementation of TQF and iTQF models is the crucial next step in the
exploration of this interesting physics.

 {\it Note added.} Recently, the authors (together with L. Pollet) performed a numerically exact simulation of the lattice realization of model D \cite{PRAus}. 
 The results are in perfect agreement with the effective field theory of the present paper. This, together with the potential experimental realizability and interest from a number of leading AMO groups in simulating model D experimentally provides significant motivation to our study.

{\it Acknowledgments}. The authors are grateful to Dam Son for a
discussion of their results and in particular, bringing attention to
Ref.~\cite{KunYangEdge}. We are also thankful to Yutaka Sasaki for
discussing the domain walls in liquid $^3$He.  We are grateful to an anonymous referee for
pointing out to us that the iTQF physics also takes place in the setups of Refs.~\cite{Pankov,Werner,Caza,Lobo}. This work was supported
by the National Science Foundation under the grants DMR-2032136,
DMR-2032077,  DMR-2335905 and DMR-2335904.  LR acknowledges support by the Simons Investigator Award
from the Simons Foundation, and thanks Kavli Institute for Theoretical
Physics for its hospitality, supported in part by the National Science
Foundation under Grant No. PHY-2309135.

\end{document}